\documentclass[a4paper, twocolumn, superscriptaddress, nofootinbib, showpacs,
amsmath, amssymb]{revtex4}

\begin{document}

\title{ Corrections to the black body radiation due to minimum-length \\deformed quantum mechanics }

\author{  Data~Mania }
\email{mania@phys.ksu.edu} \affiliation{Department of Physics, Kansas State University, 116 Cardwell Hall, Manhattan, KS 66506, USA} \affiliation{ Center for Elementary Particle Physics, ITP, Ilia State University, 3/5 Cholokashvili Ave., Tbilisi 0162, Georgia }

\author{ Michael~Maziashvili}
\email{maziashvili@gmail.com} \affiliation{ Center for Elementary Particle Physics, ITP, Ilia State University, 3/5 Cholokashvili Ave., Tbilisi 0162, Georgia }

\begin{abstract}

Planck spectrum of black body radiation is usually derived by considering of quantized free electromagnetic field at a finite temperature. The minimum-length deformed quantization affects field theory both at the first and second quantization levels. Performing an exact calculation to the first order in deformation parameter, both of the corrections turn out to be of the same order. Nevertheless, the correction at the second quantization level has some qualitative difference, that may be interesting for future study to differentiate between these two sorts of corrections. In itself the correction to the black body radiation seems to be innocuous in light of the big-bang nucleosynthesis whenever the minimum length is less or equal to $10^{-19}$cm.

\end{abstract}

\pacs{04.60.Bc }


\maketitle

\section{Introduction  }
\label{Introduction}

The minimum-length deformed quantum mechanics results in generalized uncertainty relation implying the existence of a minimum length below which the uncertainty in position cannot be reduced \cite{KMM}. The generalized uncertainty relation has appeared in the context of perturbative string theory as a consequence of the fact that strings cannot probe distances below the string scale (string length) \cite{String}. Further motivation for this generalized uncertainty relation was found in the framework of {\tt Gedankenexperiments} combining quantum theory and general relativity \cite{heuristic}.

In what follows we will assume system of units $\hbar = c = k_B =
1$. The minimum-length modified commutation relation
has the form \cite{KMM}

\begin{equation}\label{minlengthqm} [\hat{X}_i,\hat{P}_j]=i
\left(\delta_{ij}+\beta\hat{P}^2\delta_{ij} +
\beta^\prime\hat{P}_i\hat{P}_j \right)~.\end{equation}
This modified quantum mechanics imposes lower bound on position
uncertainty $\delta x$
\[\delta x_{min}=\sqrt{3\beta+\beta^\prime}~,\] where the minimum length is usually set by the quantum
gravity scale $\delta x_{min}= l_P\simeq 10^{-33}$\,cm or by the string length in string-gravity framework which is as small as $l_P$ (string length is defined by means of the string tension $\sigma$ (i.e.,
energy per unit of length) as $l_S  = \sigma^{-1/2}$). In a particular case
$\beta^\prime=2\beta$, i.e.,
\begin{equation}\label{partcase}
[ \hat{X}_i, \hat{P}_j ] = i ( \delta_{ij}
          + \beta \hat{P}^2 \delta_{ij}
          + 2\beta \hat{P}_i \hat{P}_j
        )~, \nonumber \end{equation} the realization of this algebra to the linear order in
$\beta$ can be done in a simple way in terms of the standard
position and momentum operators $ \left[ \hat{x}_i,\, \hat{p}_j
\right] = i \delta_{ij}$ \cite{Brau}
\begin{equation}\label{brauapproach} \hat{X}_i = \hat{x}_i~,~~
~~ \hat{ P}_i = \hat{ p}_i \left[ 1 + {\beta} \left(\hat{\mathbf
p}\right)^2 \right] ~.\end{equation} In the classical limit the
Eq.(\ref{brauapproach}) admits simple physical interpretation that
due to quantum-gravitational fluctuations of the background metric the momentum
${\mathbf p}$ acquires the increment ${\beta} {\mathbf
p}^2{\mathbf p}$ yielding the modified dispersion relation 

\begin{equation}  \varepsilon^2 = {\mathbf p}^2 + m^2 + 2{\beta} {\mathbf p}^4~. \label{moddisprelation}\end{equation}

In this paper we analyse the corrections to the free quantum field theory due to minimum-length deformed quantum mechanics and subsequent consequences for the black body radiation. The effect of modified dispersion relation on black body radiation have been studied in much detail \cite{Niemeyer, Das:2010gk, Zhang:2011ms}. Let us notice one more paper \cite{ChMOT} where the classical limit of minimum-length deformed quantum mechanics and its consequences for black body radiation has been studied. The new point that is the main focus of our paper is related to the field quantization with respect to the minimum-length deformed quantum mechanics.

Albeit there are several approaches to quantum gravity, one may hope that the generalized uncertainty relation suggested in \cite{String}, and a particular construction of the deformed quantum mechanics based on it, Eq.(\ref{partcase}), catches certain features of quantum gravity.

\section{ Quantum field theory in light of the minimum-length deformed quantum mechanics }
\label{ML_QFT}

The departure from usual quantum field theory (QFT) takes place in two ways. First, the modified dispersion relation affects the classical field theory. Second, the modified quantum mechanics will come into play when we are quantizing the field.

\subsection{ Imprints of modified dispersion relation on a classical field }
\label{ML_CF}

Usually, in constructing of relativistic quantum mechanics the energy and momentum
operators $\widehat{ E} = -i\partial_t,\,\widehat{\mathbf p} =
-i\partial_{\mathbf r}$ are setting the stage through the
energy-momentum four-vector operator $\widehat{ p}_{\mu} =
-i\partial_{\mu}$ \cite{FG}. Following this prescription one can
write the action as \cite{Kempf}

\begin{equation}\label{scaction}  \mathcal{A}[\Phi] = - \int d^4x \left[{\Phi\partial_t^2\Phi \over 2} + {\Phi\hat{\mathbf P}^2\Phi \over 2}  + U(\Phi)\right]~,\end{equation} that results in the equation of motion

\[ \partial_t^2\Phi +  \hat{\mathbf P}^2\Phi + U'(\Phi)  = 0  ~,
\] where the prime denotes differentiation with respect to $\Phi$. Substituting the Eq.(\ref{brauapproach}), the modified equation of motion to the lowest order in $\beta$ 
takes the form \[ \partial_t^2\Phi -
\Delta\Phi + 2\beta \Delta\Delta\Phi + U'(\Phi) = 0 ~.
\] The energy functional looks now like
\begin{eqnarray}&& E[\Phi]  = \nonumber \\ && \int d^3x \left[{(\partial_t\Phi)^2 \over 2}  +  \beta\partial_i\partial_k\Phi \,\partial_i\partial_k\Phi
  +  {\partial_i\Phi\partial_i\Phi \over 2}
 + U(\Phi)\right] ~,\nonumber \end{eqnarray} where the summation over $i,\,k$
 indices are implied.

\subsection{ The effect of modified quantum mechanics on a quantum field } 
\label{ML_QF}

We set $U(\Phi) = 0$ as for our further purposes it does not play any role. Putting the field in a box with sides of length $l$, imposing periodic boundary conditions, $\Phi(x^i + l) = \Phi(x^i)$, where $i = 1,\,2,\,3,$ and using the Fourier expansion \[ \Phi(\mathbf r) = {1 \over
\sqrt{l^3}}\sum\limits_{{\mathbf k}_n} \varphi({\mathbf
k}_n)\,e^{-i{\mathbf k}_n {\mathbf r}
}\,,~~~{\mathbf k}_n \equiv {2\pi \over l}(n_1,\,
n_2,\,n_3)~,\] from Eq.(\ref{scaction}) one finds \begin{equation}\label{modac} \mathcal{A}[\Phi] =
  \sum\limits_{{\mathbf k}_n \geq 0} \, \int\limits_{t_1}\limits^{t_2}dt
 \left[ \,\left|\dot{\varphi}({\mathbf
k}_n)\right|^2 - \left({\mathbf k}_n^2 + 2\beta {\mathbf k}_n^4 \right) \left|\varphi({\mathbf
k}_n)\right|^2  \right]~. \end{equation} We have zero mode, ${\mathbf k}_n = {\mathbf 0}$, plus oscillators
which we have to quantize with respect to the modified quantum mechanics. This is the second related departure from usual theory.

\section{ Dynamical system in thermal equilibrium: Harmonic oscillator and black body radiation }

Assume that the Hamiltonian $\widehat{\mathcal{H}}$, the
eigenvalues $E_n$ and the eigenstates $| n \rangle$ of the system
are known. In thermal equilibrium, the probability of finding the
system in the $n$-th state at the temperature $T$ is given by the Boltzmann distribution \cite{Boltzmann}

\begin{equation}\label{Boltzmann}W_n = { e^{-E_n/T} \over  \sum\limits_{n}  e^{-E_n/T} } ~.\end{equation}
The mean values of energy and pressure at a temperature $T$ can be simply
estimated by means of this expression

\begin{equation}\label{meanenpre} E = \sum\limits_{n} E_n W_n~,~~~~ P = -\sum\limits_{n}  W_n {\partial E_n
\over \partial V}~. \end{equation} The harmonic oscillator

\[
\hat{\mathcal{H}} = {\hat{p}^2 \over 2m}
        +  {m\,\omega^2
\hat{q}^2 \over 2}~,
\] has well-known energy spectrum \[ E_n = \omega\left(n + {1 \over
2}\right)~, \] for which one gets

\[ \sum\limits_{n}  e^{-\omega\left.\left(n
+ {1 \over 2}\right)\right/T}
 = { e^{-\omega/2T} \over 1 -  e^{- \omega/T}}~,~ \Rightarrow ~~W_n  =  { 1 -  e^{-\omega/T} \over e^{\omega n/T}}
 ~.\] Using the Eq.(\ref{meanenpre}), the mean value of energy at temperature $T$ takes the form

\begin{equation}\label{meanenosconedim} E = { \omega \over 2} + { \omega  \over  e^{\omega/T} - 1} ~.\end{equation}

Now let us turn to field theory. The free electromagnetic field enclosed into the box $l^3$ is usually
represented (upon making the Fourier transform) as a sum of (one-dimensional) harmonic oscillators \cite{LL}

\begin{equation}\label{emham} \hat{\mathcal{H}}  = {1 \over 2}  \sum\limits_{\alpha,\,{\mathbf k}_n \geq 0}   \left[\, l_{\star}\hat{p}_{\alpha,\,{\mathbf k}_n}^2 + {{\mathbf
k}_n^2 \over l_{\star}}\, \hat{q}_{\alpha,\,{\mathbf k}_n}^2  \right] ~,\end{equation}
where the length scale $l_{\star}$ may in fact be arbitrary,   
\begin{equation}\label{wavenumb}{\mathbf k}_n \equiv {2\pi \over l}(n_1,\,
n_2,\,n_3)~~\mbox{and}~~\alpha = 1,\,2~. 
\end{equation}

Let us digress a moment to point out the meaning and importance of length scale $l_{\star}$. The point is that in the standard quantization the energy spectrum of harmonic oscillator does not depend on mass, so in the standard case the explicit value of the length scale $l_{\star}$ is of little interest. But a remarkable feature of minimum-length deformed quantization is that the energy spectrum of harmonic oscillator becomes mass dependent, see Eq.(\ref{mlcorosc}) below. Lets us notice that the only length scale we have in the case of a free, massless QFT at a finite temperature is $T^{-1}$. So it is straightforward to identify the length scale $l_{\star}$ for the problem under consideration.

Each term in the sum (\ref{emham}) corresponds to a definite wave vector, ${\mathbf k}_n$, and polarization $\alpha$.
Knowing the energy spectrum of harmonic oscillator we can
therefore immediately write down the field mean energy at a temperature $T$

\begin{equation}\label{elmagnfielden} E = 2\sum\limits_{{\mathbf k}_n }\left[{ \omega_{{\mathbf k}_n} \over 2} + { \omega_{{\mathbf k}_n} \over  e^{\omega_{{\mathbf k}_n}/T} -
1}\right] ~,\end{equation} where the pre-factor $2$ accounts for the
polarization degrees of freedom and $\omega_{{\mathbf k}_n} =
\sqrt{{\mathbf k}_n^2}$. Defining the occupation number

\[ \bar{n}_{{\mathbf k}_n} = {1 \over e^{ \omega_{{\mathbf k}_n}/T} - 1}~,\] and omitting the zero-point energy, the Eq.(\ref{elmagnfielden}) takes the form 

\[ E = 2 \sum\limits_{{{\mathbf k}_n}} {\omega_{{\mathbf k}_n} \over e^{ \omega_{{\mathbf k}_n}/T} - 1}   = 2 \sum\limits_{{{\mathbf k}_n}}\bar{n}_{{\mathbf k}_n} \omega_{{\mathbf k}_n} ~.
\]
 If $l$ is sufficiently large then the momentum spacing $2\pi / l$
 becomes small and we can replace this sum by the integral. The total number of possible values of the
 vector ${\mathbf k}_n$ in the interval ${\mathbf k},\, {\mathbf k} + d{\mathbf k}$ is equal to
$d^3k\,l^3/(2\pi)^3$, i.\,e., for a spherical layer $4\pi
k^2dk\,l^3/(2\pi)^3$. In this case the energy density takes the
form

\[ \rho = {E \over l^3} =   {8\pi \over (2\pi)^3} \int\limits_{0}\limits^{\infty} {\omega^3d\omega \over  e^{\omega /T} - 1}
= {\pi^2T^4 \over 15\,}~.\] One easily finds the pressure by observing that energy levels of Hamiltonian (\ref{emham}) depend on the volume through the wave vectors ${\mathbf k}_n=V^{-1/3} 2\pi(n_1,n_2,n_3)$; so the factor $\partial E_N/\partial V$ entering the expression of pressure in Eq.(\ref{meanenpre}) takes the form $\partial E_n/\partial V = - E_n/3V$ and respectively for the pressure one finds 

\[P =  {1 \over 3V}\sum\limits_{n} E_n W_n = {\rho \over 3} ~.\]

\section{\textsf{  Black body radiation corrected due to minimum-length deformed quantum mechanics  }}

\subsection{Harmonic oscillator}

The lowest
order correction to the energy spectrum of (one-dimensional) harmonic oscillator due to minimum-length modified quantum mechanics has the following form \cite{MLoscillator} \begin{equation}\label{mlcorosc} \widetilde{E}_n = \omega \left( n + {1 \over 2}\right) +
m\omega^2\beta \left({n^2 \over 2} + {n \over 2} + {1 \over 4}
\right) ~.\end{equation} The Boltzmann distribution, Eq.(\ref{Boltzmann}), in the first approximation takes the form   

\begin{widetext}
\begin{eqnarray} \widetilde{W}_n \,=\, \frac{e^{-(E_n +\delta E_n)/T}}{\sum\limits_n e^{-(E_n +\delta E_n)/T}} &=& \left(e^{-E_n/T}  \,-\,e^{-E_n/T} \,\frac{\delta E_n}{T}\right)\left( \frac{1}{\sum\limits_n e^{-E_n/T}} \,+\, \frac{\sum\limits_n e^{-E_n/T} \delta E_n}{T\left[\sum\limits_n e^{-E_n/T} \right]^2} \right) \,=\, \nonumber \\&& W_n \,-\, \frac{\delta E_n}{T}\,W_n \,+\, \frac{\delta E}{T}\,W_n ~,  \end{eqnarray}\end{widetext} where $W_n$ denote zero-order values and $\delta E$ stands for mean value of energy correction \[ \delta E \,=\, \sum\limits_n \delta E_n W_n ~.\] For mean energy, in the first approximation one finds the following expression

\begin{eqnarray}\label{onedimcorrmeanenosc}&& \sum\limits_n \left(E_n +\delta E_n \right) \widetilde{W}_n \,=\,\nonumber \\&& E \,+\, \delta E  \,-\, \sum\limits_n \frac{\delta E_n}{T}\,E_nW_n \,+\, \frac{\delta E}{T}\,E~, \end{eqnarray} where all averages are taken with respect to the $W_n$. Using the relations 

\newpage
\begin{widetext}
\begin{eqnarray}  \sum\limits_n n W_n \,=\,  \frac{e^{-\omega /T}}{1-e^{-\omega /T}} ~,~~~ \sum\limits_n n^2 W_n \,=\,  \frac{e^{-\omega /T} +e^{-2\omega /T}}{ \left( 1-e^{-\omega /T}\right)^2 }~, ~~~ \sum\limits_n n^3 W_n \,=\, \frac{e^{-\omega /T} +4e^{-2\omega /T} + e^{-3\omega /T}}{ \left( 1-e^{-\omega /T}\right)^3 }~,\nonumber \end{eqnarray} one finds

\begin{eqnarray}\label{aux1} && \delta E \,=\, \frac{ m\omega^2\beta}{2}\sum\limits_n   \left(n^2  + n  + {1 \over 2}
\right) W_n \,=\,   \frac{m\omega^2\beta}{4} +  \frac{m\omega^2\beta}{2} \, \frac{2 e^{-\omega /T}}{\left(1-e^{-\omega /T}\right)^2} ~, \\\label{aux2} && \sum\limits_n \delta E_nE_nW_n \,=\, \frac{\omega}{2}\sum\limits_n \delta E_nW_n + \omega \sum\limits_n n \delta E_nW_n \,=\,  \frac{\omega}{2}\, \delta E +  \frac{m\omega^3\beta}{4} \,\, \frac{5e^{-\omega /T} + 6 e^{-2\omega /T} + e^{-3\omega /T}}{\left(1-e^{-\omega /T}\right)^3}~.  \end{eqnarray}

\noindent  By using Eqs.(\ref{meanenosconedim}, \ref{aux1}, \ref{aux2}), for Eq.(\ref{onedimcorrmeanenosc}) one gets

\begin{eqnarray}&&  \sum\limits_n \left(E_n +\delta E_n \right) \widetilde{W}_n \,=\, E \,+\, \delta E \,+\, \frac{\delta E}{T} \,\frac{\omega e^{-\omega /T}}{1-e^{-\omega /T}} \,-\,  \frac{m\omega^3\beta}{4 T} \,\, \frac{5e^{-\omega /T} + 6 e^{-2\omega /T} + e^{-3\omega /T}}{\left(1-e^{-\omega /T}\right)^3} \,=  \nonumber \\&& \frac{\omega}{2}+ \frac{m\omega^2\beta}{4} + \frac{\omega e^{-\omega /T}}{1-e^{-\omega /T}} + m\omega^2\beta \, \frac{e^{-\omega /T}}{\left(1-e^{-\omega /T}\right)^2} + \left[\frac{m\omega^2\beta}{4 T} +  \frac{m\omega^2\beta}{T} \, \frac{ e^{-\omega /T}}{\left(1-e^{-\omega /T}\right)^2} \right] \,\frac{\omega e^{-\omega /T}}{1-e^{-\omega /T}}  \,- \nonumber \\ &&  \frac{m\omega^3\beta}{4 T} \,\, \frac{5e^{-\omega /T} + 6 e^{-2\omega /T} + e^{-3\omega /T}}{\left(1-e^{-\omega /T}\right)^3}  \,=\, \nonumber \\&&  \frac{\omega}{2} \,+\, \frac{m\omega^2\beta}{4} \,+\, \frac{\omega e^{-\omega /T}}{1-e^{-\omega /T}} \,+\, m\omega^2\beta \, \frac{e^{-\omega /T}}{\left(1-e^{-\omega /T}\right)^2}  \,-\, \frac{m\omega^3\beta}{4 T} \,\, \frac{e^{-\omega /T} + 10 e^{-2\omega /T} -3 e^{-3\omega /T}}{\left(1-e^{-\omega /T}\right)^3}  ~. \label{odosccorrmeanen}\end{eqnarray}\end{widetext}

\subsection{Turning to the field theory} 
 
Turning now to the field theory, from Eq.(\ref{modac}) one finds that the Hamiltonian of free electromagnetic field, Eq.(\ref{emham}), takes the form 

\begin{equation}\label{modemham} \hat{\mathcal{H}}  = {1 \over 2}  \sum\limits_{\alpha,\,{\mathbf k}_n \geq 0}   \left[\, l_{\star} \,\hat{p}_{\alpha,\,{\mathbf k}_n}^2 + {{\mathbf
k}_n^2 + 2\beta {\mathbf k}_n^4 \over l_{\star}  }\, \hat{q}_{\alpha,\,{\mathbf k}_n}^2  \right] ~.\end{equation} Upon quantizing with respect to the minimum-length deformed quantum mechanics each oscillator entering the Hamiltonian (\ref{modemham}) results in

 \begin{equation}\label{mlcorharmonics} \widetilde{E}_{{\mathbf k}_n} = \left(k_n + \beta k_n^3 \right) \left( N + {1 \over 2}\right) +
{\beta k_n^2 \over l_{\star}}\left({N^2 \over 2} + {N \over 2} + {1 \over 4}
\right) ~.\end{equation} 

\noindent Now for estimating of energy density we have to drop zero-point energy in Eq.(\ref{odosccorrmeanen}), make the replacements $m \rightarrow l_{\star}^{-1}\,,\, \omega \rightarrow \omega_{{\mathbf k}_n} + \beta \omega_{{\mathbf k}_n}^3$ and take the sum over all ${\mathbf k}_n$ modes (see the Appendix)

\begin{widetext}
\begin{eqnarray}&&  \rho \,=\, \frac{2}{l^3} \int\limits_{0}^{\infty} \frac{4\pi l^3
\omega^2d\omega}{(2\pi)^3} \left[ \frac{\left(\omega +\beta\omega^3\right) e^{-\left(\omega +\beta\omega^3\right) /T}}{1-e^{-\left(\omega +\beta\omega^3\right) /T}} \,+\, \frac{\beta\omega^2}{l_{\star}} \, \frac{e^{-\omega /T}}{\left(1-e^{-\omega /T}\right)^2}  \,-\, \frac{\beta\omega^3}{4 l_{\star}T} \,\, \frac{e^{-\omega /T} + 10 e^{-2\omega /T} -3 e^{-3\omega /T}}{\left(1-e^{-\omega /T}\right)^3}  \right] \,= \nonumber \\ &&  \frac{\pi^2}{15}\,T^4  \,-\, \frac{40\pi^4}{63}\,\beta T^6  \,-\,  \frac{\left[74\zeta(4)\Gamma(4) \,-\, 3\zeta(6)\Gamma(6) \right]}{4 l_{\star}\pi^2}\, \beta T^5  \,=\, \frac{\pi^2}{15}\,T^4  \,-\, \frac{40\pi^4}{63}\,\beta T^6  \,-\,  \left( \frac{74\pi^2}{15} \,-\, \frac{24\pi^4}{63} \right)\frac{\beta T^5}{4 l_{\star}} ~. \label{compenensitytofisrtbeta}\end{eqnarray}

For finding pressure one notices that energy levels $\widetilde{E}_{{\mathbf k}_n}$ in Eq.(\ref{mlcorharmonics}) depend on the volume through the wave vectors ${\mathbf k}_n=V^{-1/3} 2\pi(n_1,n_2,n_3)$; so the factor $\partial \widetilde{E}_{{\mathbf k}_n}/\partial V$ entering the expression of pressure in Eq.(\ref{meanenpre}) takes the form 

\[\frac{\partial \widetilde{E}_{{\mathbf k}_n}}{\partial V}\,=\, -\frac{1}{3V}\left[\left(k_n + 3 \beta k_n^3 \right) \left( N + {1 \over 2}\right) + 
{2\beta k_n^2 \over l_{\star}}\left({N^2 \over 2} + {N \over 2} + {1 \over 4}
\right)\right] ~.  \] Let us first focus on the correction coming from the $l_{\star}$ term. The modification as compared to the above calculations is that the term given by Eq.(\ref{aux1}) will enter the pressure with the additional factor $2/3$. So, now we will have 

\begin{eqnarray}&&  \frac{1}{\pi^2} \int\limits_{0}^{\infty} 
\omega^2d\omega  \left[  \frac{2\beta\omega^2}{3l_{\star}} \, \frac{e^{-\omega /T}}{\left(1-e^{-\omega /T}\right)^2}  \,-\, \frac{\beta\omega^3}{4 l_{\star}T} \,\, \frac{e^{-\omega /T} + 10 e^{-2\omega /T} -3 e^{-3\omega /T}}{\left(1-e^{-\omega /T}\right)^3}  \right] \,= \nonumber \\ &&\frac{2}{3}\,  \frac{16\zeta(4)\Gamma(4) }{4 l_{\star}\pi^2}\, \beta T^5  \,-\,  \frac{\left[80\zeta(4)\Gamma(4) \,-\, 3\zeta(6)\Gamma(6) \right]}{4 l_{\star}\pi^2}\, \beta T^5  \,=\,  \,-\,\left(\frac{208\pi^2}{45} \,-\, \frac{24\pi^4}{63} \right)  \frac{\beta T^5 }{4 l_{\star}}  ~.\end{eqnarray} The remaining terms to the first order in $\beta$ can be estimated straightforwardly

\begin{eqnarray}&&\int\limits_0^\infty \omega^2d\omega \,\frac{\omega/3 + \beta \omega^3}
{\exp\left(\frac{\omega \,+\, \beta \omega^3 }{T}\right)-1} \,=\,
\int\limits_0^\infty \omega^2d\omega \left(\omega/3+\beta \omega^3\right)\left(\frac{1}{e^{\omega/T}-1}- \beta\, \frac{\omega^3 e^{\omega/T} }{T(e^{\omega/T}-1)^2}+O\left(\beta^2\right)\right) \,
=  \nonumber\\
&&\frac{1}{3}\int\limits_0^\infty \omega^3d\omega\frac{1}{e^{\omega/T}-1}
-\frac{\beta}{3T}\int\limits_0^\infty \omega^6d\omega\frac{e^{\omega/T}}{(e^{\omega/T}-1)^2}
+\beta\int\limits_0^\infty \omega^5d\omega\frac{1}{e^{\omega/T}-1}
 \,
=  \frac{\zeta(4)\,\Gamma(4)}{3}\,T^4 \,-\, \beta\,\zeta(6)\,\Gamma(6)\,T^6 \,=\,\nonumber\\&&  \frac{\pi^4}{45}\,T^4 \,-\, \frac{8\pi^6}{63}\,\beta T^6 ~. \nonumber
\end{eqnarray}\end{widetext}

\noindent Thus the final result takes the form 

\begin{equation}
P \,=\, \frac{\pi^2}{45}\,T^4 \,-\, \frac{8\pi^4}{63}\,\beta T^6  \,-\,\left(\frac{208\pi^2}{45} \,-\, \frac{24\pi^4}{63} \right)  \frac{\beta T^5 }{4 l_{\star}}  ~. \label{compprestofisrtbeta}\end{equation}

Now let us specify the length scale $l_{\star}$. The natural characteristic energy scale for the problem under consideration is the temperature $T$. Moreover, that is the only energy scale at hand. So, to fix $l_{\star}$, we have to identify $l_{\star} = T^{-1}$. Then the equations (\ref{compenensitytofisrtbeta},\, \ref{compprestofisrtbeta}) take the form

\begin{eqnarray}\label{feqendensity}
&& \rho \,=\, \frac{\pi^2}{15}\,T^4    \,-\,  \left( \frac{18.5\pi^2}{15} \,+\, \frac{34\pi^4}{63} \right) \, \beta T^6 ~, \\&& P  \,=\, \frac{\pi^2}{45}\,T^4 \,-\,\left(\frac{54\pi^2}{45} \,+\, \frac{2\pi^4}{63} \right)  \, \beta T^6~. \label{feqpres}
\end{eqnarray}

\noindent Entropy density gets the following correction 

\begin{equation}\label{entropydensity} s \,=\, \frac{\rho +P}{T} \,=\, \frac{4\pi^2}{45}\,T^3   \,-\,  \left( \frac{109.5\pi^2}{15} \,+\, \frac{36\pi^4}{63} \right) \, \beta T^5 ~.  \end{equation}

\noindent The validity condition for using the Eqs.(\ref{feqendensity}, \ref{feqpres}, \ref{entropydensity}) is set by the requirement the $\beta$ terms to be much smaller than the leading term 

\[ \beta T^2 \ll 1. \]

\section{ Cosmological constraints on $\beta$ in light of Eqs.(\ref{feqendensity}, \ref{feqpres})}

The expansion rate of the universe is determined by the Friedmann and energy conservation equations

\begin{equation}\label{friedmannmatcons} H^2 \equiv \left( {\dot{a} \over a} \right)^2  = {8\pi \over 3m_P^2}\,\rho~, ~~~~ ~~~\dot{\rho} +3H(\rho + P ) = 0~.  \end{equation} In view of Eqs.(\ref{feqendensity}, \ref{feqpres}) the radiation energy density and pressure encompassing all relativistic particle species take the form

\begin{eqnarray}\label{endenpre}&& \rho =  g_*(T)\left[ \frac{\pi^2}{30}\,T^4    \,-\,  \left( \frac{18.5\pi^2}{30} \,+\, \frac{17\pi^4}{63} \right) \, \beta T^6 \right] ~,~~~~ \\ && P =  g_*(T)\left[  \frac{\pi^2}{90}\,T^4 \,-\,\left(\frac{27\pi^2}{45} \,+\, \frac{\pi^4}{63} \right)  \, \beta T^6 \right]  ~, \nonumber \end{eqnarray} where $g_*(T)$ as usual denotes effective number of relativistic species at temperature $T$. The factor $g_*(T)$ increases with increasing the temperature but it varies relatively slowly over the range $g_*(T < 1\,\mbox{MeV}) \simeq 4,\, g_*(1\,\mbox{MeV} < T < 300\,\mbox{MeV}) \simeq 10,\,\,g_*(300\,\mbox{MeV} < T ) \simeq 100$  \cite{Schwarz}.

It is convenient to write the Eq.(\ref{friedmannmatcons}) in the form  

\[ \left( {\dot{a} \over a} \right)^2  = {8\pi \over 3m_P^2}\rho_0\left( 1 + {\delta\rho \over \rho_0} \right) ~,\] where $\rho_0 \equiv  g_*(T) \pi^2T^4/30$ and $\delta\rho$ denotes the correction $\beta$ term in Eq.(\ref{endenpre}). The big bang nucleosynthesis is very sensitive to the expansion rate of the Universe at temperatures of the order of a few MeV, at this temperature the ratio $\delta\rho/\rho_0$ is of the order of $\sim \beta \left(1\,\mbox{MeV}\right)^2 $. By taking $\sqrt{\beta} \simeq 10^{14}l_P$ one finds $\beta \left(1\,\mbox{MeV}\right)^2 \simeq 10^{-3}$. The limitations on $\delta\rho/\rho_0$ due to big bang nucleosynthesis are of the order $\sim 10^{-2}$ \cite{BBN}.

 It is worth noticing that the limitation on $\beta$ coming from the inflationary perturbations spectrum is more severe \cite{Ashoorioon:2004vm, Ashoorioon:2004wd, Ashoorioon:2005ep, Kempf:2006wp,  Palma:2008tx}. Namely, estimating the tensor metric
perturbations and imposing the observational bound for gravitational
waves one gets $\beta H^2 \lesssim 10^{-4}$, where $H$ denotes Hubble parameter during the inflation.

\section{ Concluding remarks  }

\noindent  First let us notice that (qualitatively) Eqs.(\ref{feqendensity}, \ref{feqpres}, \ref{entropydensity}) can be viewed as obtained from the standard expressions by the replacement $T \rightarrow T - 25\beta T^3$. This generic feature was clearly emphasized in the recent papers devoted to the corrections to black body radiation due to modified dispersion relations \cite{Das:2010gk, Zhang:2011ms}. The main point of our study was to distinct corrections arising due to minimum-length deformed quantization of an electromagnetic field. In general there are two types of corrections due to minimum-length deformed quantum mechanics, first quantization corrections and second quantization corrections, respectively, which should be contrasted. In the case of a black body radiation both of these corrections appear to be of the same magnitude and sign. But in general these corrections are rather unlike. Let us touch this question in a bit more detail. A little care is needed in extending of minimum-length deformed formalism, Eq.(\ref{minlengthqm}), to the second quantization level. Looking at field theory as a limit of a discrete system (a straightforward and natural way for doing so is to represent spatial integral as a Riemann-sum over the discrete set of points) 

\[ L = \int d^3 x \mathcal{L}(\Phi,\, \dot{\Phi},\, \boldsymbol{\bigtriangledown}\Phi ) \sim \sum\limits_{\mathbf{x}_i} (\delta x)^3 \mathcal{L}(\Phi_i,\, \dot{\Phi}_i,\, [\boldsymbol{\bigtriangledown}\Phi]_i )~,\] one finds \[ p_i = \frac{\partial L}{\partial \dot{\Phi}_i} = \Pi(\mathbf{x}_i)(\delta x)^3~,\] and the Eq.(\ref{minlengthqm})  

\[ [\Phi_i,\,p_j] = i\left( \delta_{ij} + \beta p^2\delta_{ij} + \beta^\prime p_i p_j \right)~, \] 
takes the form 

\begin{widetext}
\[ \left[\Phi_i,\,\Pi(\mathbf{x}_j)\right] = i\left( \frac{\delta_{ij}}{(\delta x)^3} + \beta (\delta x)^3  \frac{\delta_{ij}}{(\delta x)^3} \sum\limits_{k} (\delta x)^3 \left[\Pi(\mathbf{x}_k)\Pi(\mathbf{x}_k)\right] + \beta^\prime (\delta x)^3 \Pi(\mathbf{x}_i)\Pi(\mathbf{x}_j) \right) ~.\] In the limit $(\delta x)^3 \rightarrow 0$ one finds   

\[ \left[\Phi(\mathbf{x}),\,\Pi(\mathbf{x}')\right] = i\left( \delta(\mathbf{x} - \mathbf{x}') + \lim\limits_{(\delta x)^3\rightarrow 0} \left[\beta (\delta x)^3   \delta(\mathbf{x} - \mathbf{x}') \int d^3y \,\Pi^2(\mathbf{y}) + \beta^\prime (\delta x)^3 \Pi(\mathbf{x})\Pi(\mathbf{x}') \right]\right) = i\delta(\mathbf{x} - \mathbf{x}') ~,\]\end{widetext} that is, the corrections to the commutator disappear. So, at the first glance it may seem that minimum-length deformed formalism does not affect the field quantization. But the oscillator expansion of the field, see Eq.(\ref{modac}), shows an obvious way for applying of minimum-length deformed formalism at the second quantization level. That is the way used in a foregoing discussion. Following this way one can estimate corrections to the black hole radiation as well \cite{Berger:2010pj}. Remarkably enough, the corrections to the black hole entropy appear to be in complete agreement with the results obtained previously in the framework of loop quantum gravity and in a tunneling formalism approach to the black hole emission. Let us notice that, one can easily expand the present analyses to a system of coupled harmonic oscillators, which naturally appears after putting the field in a spherical box and expanding it in terms of the spherical harmonics, and try to estimate the corrections to the entanglement entropy.

\vspace{0.5cm}

\noindent \vspace{0.06cm}
{\tt \bf  Acknowledgments:}  The work was supported in part by the CRDF/GRDF grant and SNSF -- SCOPES grant No. 128040. Authors are indebted to the Referee for an idea to indicate besides the bound on $\beta$ coming from nucleosynthesis the inflationary bounds as well.

\appendix

\section*{Appendix}

\begin{widetext}
To simplify the reading of the paper, we quote some basic integrals that enter the corrections to the energy density and pressure. By using Eq.(\ref{mlcorharmonics}) (without $l_{\star}$ terms) and the integral

\[ \int\limits_0^\infty { x^{n-1} \over e^x-1}dx=\zeta(n)\Gamma(n) ~,\] to the first order in $\beta$ one finds 

\begin{eqnarray}
&&\int\limits_0^\infty \omega^2d\omega \, {\omega+\beta\omega^3  \over \exp\left({\omega \,+\,\beta \omega^3  \over T}\right)-1} \,=\,   
\int\limits_0^\infty \omega^2d\omega \left(\omega+\beta \omega^3 \right)\left(\frac{1}{e^{\omega/T} - 1}\, - \,\beta\,\frac{\omega^3 e^{\omega/T}}{T\left(e^{\omega/T}-1\right)^2}+O\left(\beta^2\right)\right) \,=
\nonumber\\
&& \int\limits_0^\infty \omega^3d\omega\frac{1}{e^{\omega/T}-1}
-\frac{\beta}{T}\int\limits_0^\infty \omega^6d\omega\frac{e^{\omega/T}}{\left(e^{\omega/T} - 1\right)^2} +\beta\int\limits_0^\infty \omega^5d\omega\frac{1}{e^{\omega/T}-1} \,=\,
\zeta(4)\,\Gamma(4)\,T^4 \,-\, 5\beta\zeta(6)\,\Gamma(6)\,T^6 \,= \nonumber \\&& \frac{\pi^4}{15}T^4  \,-\, \beta\frac{40\pi^6}{63}T^6~,~~~~\mbox{where we have used}~~~\zeta(4)\Gamma(4) = \frac{\pi^4}{15}~,~~  \zeta(6)\Gamma(6) =  \frac{8\pi^6}{63}~. \nonumber
\end{eqnarray}

\noindent The other integrals we need look as follows: 

\begin{eqnarray}&& \int\limits_0^{\infty} dx \, \frac{x^4 e^{-x}}{\left(1-e^{-x}\right)^2} \,=\, \int\limits_0^{\infty} dx \, \frac{x^4 e^{x}}{\left(e^{x} -1\right)^2} \,=\, -  \int\limits_0^{\infty} x^4 \, d \frac{1}{e^{x} -1} \,=\, - \left.\frac{x^4}{e^{x} -1}\right|_{0}^{\infty} + 4  \int\limits_0^{\infty} dx \, \frac{x^3 }{e^{x} -1} \,=\,4\zeta(4)\Gamma(4) ~.\nonumber \\&&
\int\limits_0^{\infty} dx \, \frac{x^5 e^{-x}}{\left(1-e^{-x}\right)^3} \,=\, \int\limits_0^{\infty} dx \, \frac{x^5 e^{2x}}{\left(e^{x} -1\right)^3} \,=\,  -\frac{1}{2}  \int\limits_0^{\infty} x^5e^{x} \, d \frac{1}{\left(e^{x} -1\right)^2} \,=\, \nonumber \\&& -\frac{1}{2}  \left.\frac{x^5e^{x}}{\left(e^{x} -1\right)^2}\right|_0^{\infty} \,+\,\frac{1}{2}  \int\limits_0^{\infty} dx\, \frac{x^5e^{x} \,+\, 5x^4e^{x}}{\left(e^{x} -1\right)^2}  \,=\, \frac{5\zeta(5)\Gamma(5)}{2} \,+\, 10\zeta(4)\Gamma(4)~.\nonumber\\&&
\int\limits_0^{\infty} dx \, \frac{x^5 e^{-2x}}{\left(1-e^{-x}\right)^3} \,=\, \int\limits_0^{\infty} dx \, \frac{x^5 e^{x}}{\left(e^{x}-1\right)^3} \,=\, -\frac{1}{2}  \left.\frac{x^5 }{\left(e^{x}-1\right)^2}\right|_{0}^{\infty} \,-\, \frac{5}{2}\int\limits_0^{\infty} dx \, \frac{x^4 }{e^{x}-1} \,+\, \frac{5}{2}\int\limits_0^{\infty} dx \, \frac{x^4 e^{x}}{\left(e^{x}-1\right)^2} \,=\, \nonumber \\ && 10\zeta(4)\Gamma(4)  \,-\, \frac{5\zeta(5)\Gamma(5)}{2} ~. \nonumber\\ &&  \int\limits_0^{\infty} dx \, \frac{x^5 e^{-3x}}{\left(1-e^{-x}\right)^3} \,=\, \int\limits_0^{\infty} dx \, \frac{x^5 }{\left(e^{x}-1\right)^3} \,=\, \int\limits_0^{\infty} dx \, \frac{x^5 }{e^{x}-1} - \int\limits_0^{\infty} dx \, \frac{x^5e^{x} }{\left(e^{x}-1\right)^2} + \int\limits_0^{\infty} dx \, \frac{x^5 e^{x}}{\left(e^{x}-1\right)^3} \,=\, \nonumber \\&&  \zeta(6)\Gamma(6) \,-\, 5\zeta(5)\Gamma(5) \,+\, 10\zeta(4)\Gamma(4)  \,-\, \frac{5\zeta(5)\Gamma(5)}{2} ~. \nonumber
\end{eqnarray}

\end{widetext}

\end{document}